\documentclass[12pt]{iopart}

\usepackage{fancyhdr,graphicx}
\usepackage{float}
\usepackage{color}
\usepackage{iopams,wasysym}
\usepackage{cite}
\usepackage{comment}

\let\olditemize=\itemize
\def\itemize{
\olditemize
  \setlength{\itemsep}{5pt}
  \setlength{\parskip}{-5pt}
}

\begin{document}

\title[LHC Optics Measurement with Proton Tracks by TOTEM]{LHC Optics Measurement with Proton Tracks Detected by the Roman Pots of the TOTEM Experiment}

\author{
The TOTEM Collaboration:
G.~Antchev$^{\rm a}$,
P.~Aspell$^{\rm 8}$,
I.~Atanassov$^{\rm 8}$$^{\rm , a}$,
V.~Avati$^{\rm 8}$,
J.~Baechler$^{\rm 8}$,
V.~Berardi$^{\rm 5a}$$^{\rm , 5b}$,
M.~Berretti$^{\rm 7b}$,
E.~Bossini$^{\rm 7b}$,
U.~Bottigli$^{\rm 7b}$,
M.~Bozzo$^{\rm 6a}$$^{\rm , 6b}$,
E.~Br\"{u}cken$^{\rm 3a}$$^{\rm , 3b}$,
A.~Buzzo$^{\rm 6a}$,
F.~S.~Cafagna$^{\rm 5a}$,
M.~G.~Catanesi$^{\rm 5a}$,
C.~Covault$^{\rm 9}$,
M.~Csan\'{a}d$^{\rm 4}$$^{\rm , b}$,
T.~Cs\"{o}rg\H{o}$^{\rm 4}$,
M.~Deile$^{\rm 8}$,
M.~Doubek$^{\rm 1b}$,
K.~Eggert$^{\rm 9}$,
V.~Eremin$^{\rm c}$,
F.~Ferro$^{\rm 6a}$,
A. Fiergolski$^{5a}$$^{\rm , d}$,
F.~Garcia$^{\rm 3a}$,
V.~Georgiev$^{\rm 11}$,
S.~Giani$^{\rm 8}$,
L.~Grzanka$^{\rm 10}$$^{\rm , e}$,
J.~Hammerbauer$^{\rm 11}$,
J.~Heino$^{\rm 3a}$,
T.~Hilden$^{\rm 3a}$$^{\rm , 3b}$,
A.~Karev$^{\rm 8}$,
J.~Ka\v{s}par$^{\rm 1a}$$^{\rm , 8}$,
J.~Kopal$^{\rm 1a}$$^{\rm , 8}$,
V.~Kundr\'{a}t$^{\rm  1a}$,
S.~Lami$^{\rm 7a}$,
G.~Latino$^{\rm 7b}$,
R.~Lauhakangas$^{\rm 3a}$,
T.~Leszko$^{\rm  d}$,
E.~Lippmaa$^{\rm 2}$,
J.~Lippmaa$^{\rm 2}$,
M.~V.~Lokaj\'{i}\v{c}ek$^{\rm 1a}$,
L.~Losurdo$^{\rm 7b}$,
M.~Lo~Vetere$^{\rm 6a}$$^{\rm , 6b}$,
F.~Lucas~Rodr\'{i}guez$^{\rm 8}$,
M.~Macr\'{i}$^{\rm 6a}$,
T.~M\"aki$^{\rm 3a}$,
A.~Mercadante$^{\rm 5a}$,
N.~Minafra$^{\rm 5b}$$^{\rm , 8}$,
S.~Minutoli$^{\rm 6a}$,
F.~Nemes$^{\rm 4}$$^{\rm , b}$,
H.~Niewiadomski$^{\rm 8}$,
E.~Oliveri$^{\rm 7b}$,
F.~Oljemark$^{\rm 3a}$$^{\rm , 3b}$,
R.~Orava$^{\rm 3a}$$^{\rm , 3b}$,
M.~Oriunno$^{\rm f}$,
K.~\"{O}sterberg$^{\rm 3a}$$^{\rm , 3b}$,
P.~Palazzi$^{\rm 7b}$,
Z.~Peroutka$^{\rm 11}$,
J.~Proch\'{a}zka$^{\rm 1a}$,
M.~Quinto$^{\rm 5a}$$^{\rm , 5b}$,
E.~Radermacher$^{\rm 8}$,
E.~Radicioni$^{\rm 5a}$,
F.~Ravotti$^{\rm 8}$,
E.~Robutti$^{\rm 6a}$,
L.~Ropelewski$^{\rm 8}$,
G.~Ruggiero$^{\rm 8}$,
H.~Saarikko$^{\rm 3a}$$^{\rm , 3b}$,
A.~Scribano$^{\rm 7b}$,
J.~Smajek$^{\rm 8}$,
W.~Snoeys$^{\rm 8}$,
J.~Sziklai$^{\rm 4}$,
C.~Taylor$^{\rm 9}$,
N.~Turini$^{\rm 7b}$,
V.~Vacek$^{\rm 1b}$,
J.~Welti$^{\rm 3a}$$^{\rm , 3b}$,
J.~Whitmore$^{\rm g}$,
P.~Wyszkowski$^{\rm 10}$,
K.~Zielinski$^{\rm 10}$
}

\address{$^{1a}$ Institute of Physics, ASCR, Praha, Czech Republic,}
\address{$^{1b}$ Czech Technical University, Praha, Czech Republic,}
\address{$^{2\phantom{a}}$ National Institute of Chemical Physics and Biophysics NICPB, Tallinn, Estonia,}
\address{$^{3a}$ Helsinki Institute of Physics, Helsinki, Finland,}
\address{$^{3b}$ Department of Physics,  University of Helsinki, Helsinki, Finland,}
\address{$^{4\phantom{a}}$ MTA Wigner Research Center,  RMKI Budapest, Hungary,}
\address{$^{5a}$ INFN Sezione di Bari, Bari, Italy,}
\address{$^{5b}$ Dipartimento Interateneo di Fisica di  Bari, Italy,}
\address{$^{6a}$ INFN Sezione di Genova, Genova, Italy,}
\address{$^{6b}$ Universit\`{a} degli Studi di Genova,  Genova, Italy,}
\address{$^{7a}$ INFN Sezione di Pisa, Pisa, Italy,}
\address{$^{7b}$ Universit\`{a} degli Studi di Siena and Gruppo Collegato INFN di Siena,  Siena, Italy,}
\address{$^{8\phantom{a}}$ CERN, Geneva, Switzerland,}
\address{$^{9\phantom{a}}$ Case Western Reserve University,  Dept. of Physics, Cleveland, OH, USA,}
\address{$^{10}$ AGH University of Science and Technology, Krakow, Poland,}
\address{$^{11}$ University of West Bohemia, Pilsen, Czech Republic.}
\ead{frigyes.janos.nemes@cern.ch}

\begin{abstract}
Precise knowledge of the beam optics at the LHC is crucial to fulfill the
physics goals of the TOTEM experiment, where the kinematics of the scattered
protons is reconstructed with the near-beam telescopes -- so-called Roman Pots
(RP). Before being detected, the protons' trajectories are influenced by the
magnetic fields of the accelerator lattice.  Thus precise understanding of the
proton transport is of key importance for the experiment. A novel method of
optics evaluation is proposed which exploits kinematical distributions of
elastically scattered protons observed in the RPs. Theoretical predictions, as
well as Monte Carlo studies, show that the residual uncertainty of this optics
estimation method is smaller than $2.5\,\permil$.
\end{abstract}

\submitto{\NJP}
\maketitle

\section{Introduction}
\begin{figure}[!ht]
\centering
\includegraphics[trim = 0mm 0mm 0mm 0mm, clip, width=1.00\textwidth]{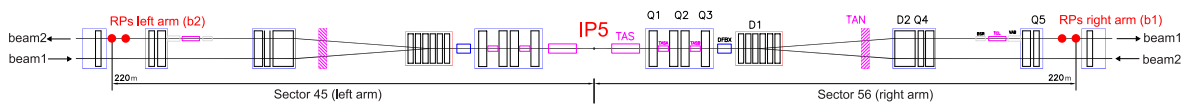}
\caption{(color online) 
Schematic layout of the LHC magnet lattice at IP5 up to the ``near'' and
``far'' Roman Pot units,
where the near and far pots are indicated by full (red) dots
on beams 1 and 2, at the positions indicated by arrows. 	 }
\label{lhc_layout_IP5}
\end{figure}

The TOTEM experiment~\cite{totem_design} at the LHC is equipped with near beam
movable insertions -- called Roman Pots (RP) -- which host silicon detectors to
detect protons scattered at the LHC Interaction Point 5
(IP5)~\cite{LHC_machine}.  This paper reports the results based on data
acquired with a total of 12 RPs installed symmetrically with respect to IP5.
Two units of 3 RPs are inserted downstream of each outgoing LHC beam: the
``near'' and the ``far'' unit located at $s=\pm214.63\,$m and $s=\pm220.00\,$m,
respectively, where $s$ denotes the distance from IP5. The arrangement of the
RP devices along the two beams is schematically illustrated in
\fref{lhc_layout_IP5}.

Each unit consists of 2 vertical, so-called ``top'' and ``bottom'', and 1
horizontal RP. The two diagonals {\it top left of IP--bottom right of IP} and
{\it bottom left of IP--top right of IP}, tagging elastic candidates, are used
as almost independent experiments. The details of the set-up are discussed
in~\cite{Anelli:2008zza}.

Each RP is equipped with a telescope of 10 silicon microstrip sensors of
$66\,\mu$m pitch which provides spatial track reconstruction resolution
$\sigma(x,y)$ of $11\,\mu$m~\cite{Antchev:2013hya}. Given the longitudinal
distance between the units of $\Delta s = 5.372\,$m the proton angles are
measured by the RPs with an uncertainty of $2.9\,\mu$rad. 

During the measurement the detectors in the vertical and horizontal RPs
overlap, which enables a precise relative alignment of all the three RPs by
correlating their positions via common particle tracks. The alignment
uncertainty better than $10\,\mu$m is attained, the details are discussed
in~\cite{Antchev:2013hya,Kaspar:Thesis}. 

The proton trajectories, thus their positions observed by RPs, are affected by
magnetic fields of the accelerator lattice. The accelerator settings define the
machine optics which can be characterized with the value of $\beta^*$ at IP5.
It determines the physics reach of the experiment~\cite{Anelli:2008zza}: runs
with high $\beta^*=90$ -- $2500\,$m are characterized by low beam divergence
allowing for precise scattering angle measurements while runs of low
$\beta^*=0.5$ -- $11\,$m, due to small interaction vertex size, provide higher
luminosity and thus are more suitable to study rare processes. In the following
sections we will analyze 
two representatives
of these LHC runs, corresponding to machine optics with 
$\beta^*=3.5\,$m and $90\,$m, respectively ~\cite{hi_beta_optics, LHC_machine}.

In order to 
reconstruct the kinematics of proton-proton scattering precisely, an accurate
model of proton transport is indispensable.  TOTEM has developed a novel method
to evaluate the optics of the machine by using angle-position distributions of
elastically scattered protons observed in the RP detectors.  The method,
discussed in detail in the following sections, has been successfully applied to
data samples recorded in 2010 and 2012~\cite{Antchev:2011zz,Antchev:2011vs,Antchev:2013gaa,Antchev:2013iaa,Antchev:2013paa}.

\Sref{proton_transport} introduces the so-called transport matrix, which
describes the proton transport through the LHC lattice, while machine
imperfections are discussed in~\sref{machine_imperfections}. The proposed novel
method for optics evaluation is based on the correlations between the transport
matrix elements. These correlations allow the estimation of those optical
functions which are strongly correlated 
to measurable combinations and estimators of certain elements of
this transport matrix. 
Therefore, it is fundamental to study these correlations in
detail, which is the subject of~\sref{corrInTranspMatrix}.  The 
corresponding eigenvector decomposition of the transport matrix 
is used to gain insight into the magnitude of the reduction of uncertainties 
in the determination of LHC optics that can be obtained from using TOTEM data
and provides the theoretical baseline of the method. 

\Sref{constrTracksRP} brings the theory to practice, by specifying the
estimators, obtained from elastic track distributions measured in RPs. Finally,
the 
algorithm that we applied to estimate the LHC optics from TOTEM data
is 
described and
discussed in~\sref{opticsMatching}.
The uncertainty of 
this novel method of LHC optics determination 
was estimated with Monte Carlo simulations,
that are 
described in detail in~\sref{monteCarloTests}. 

\section{Proton transport model}
\label{proton_transport}
Scattered protons are detected by the Roman Pots after having traversed a
segment of the LHC lattice containing 29 main and corrector magnets per beam,
shown in \fref{lhc_layout_IP5}.

The trajectory of protons produced with transverse positions\footnote{The
'$^{*}$' superscript indicates that the value is taken at the LHC Interaction
Point 5.} $(x^*,y^*)$ and angles $(\Theta_x^*, \Theta_y^*)$ at IP5 is described
approximately by a linear formula
\begin{eqnarray}
	\vec{d}(s)=T(s)\cdot\vec{d}^{*}\,,  
	\label{proton_trajectories}
\end{eqnarray} 
where $\vec{d}=\left(x,\Theta_x,y,\Theta_y,\Delta p/p\right)^{T}$, $p$ and
$\Delta p$ denote the nominal beam momentum and the proton longitudinal
momentum loss, respectively. The single pass transport matrix
\begin{eqnarray}
	T=\left(
		\begin{array}{ccccc} 
			v_x         & L_x     & m_{13}   & m_{14}  & D_x  \\
			\frac{\rmd v_x}{\rmd s}        & \frac{\rmd L_x}{\rmd s}    & m_{23}   & m_{24}  & \frac{\rmd D_x}{\rmd s} \\
			m_{31}      & m_{32}  & v_y      & L_y     & D_y  \\
			m_{41}      & m_{42}  & \frac{\rmd v_y}{\rmd s}     & \frac{\rmd L_y}{\rmd s}    & \frac{\rmd D_y}{\rmd s} \\ 
			0           & 0       & 0        & 0       & 1
    	\end{array}  
	\right)
	\label{transport_matrix}
\end{eqnarray}
is defined by the optical functions~\cite{Wiederman}. The horizontal and
vertical magnifications
\begin{equation}
v_{x,y}=\sqrt{\beta_{x,y}/\beta^*}\cos\Delta\mu_{x,y}
\end{equation}
and the effective lengths 
\begin{equation}
L_{x,y}=\sqrt{\beta_{x,y}\beta^*}\sin\Delta\mu_{x,y}
\end{equation}
are functions of the betatron amplitudes $\beta_{x,y}$ and the relative phase advance 
\begin{equation}
\Delta\mu_{x,y}=\int^{\mbox{\tiny\rm RP}}_{\mbox{\tiny\rm IP}}\frac{\rmd s}{\beta_{x,y}}\,,
\end{equation}
and are of particular importance for proton kinematics reconstruction. The
$D_x$ and $D_y$ elements are the horizontal and vertical dispersion,
respectively.

Elastically scattered protons are relatively easy to distinguish due to their scattering angle correlations. In addition, these correlations are sensitive to the
machine optics. 
Therefore, elastic proton-proton scattering measurements are 
ideally suited to investigate the optics the LHC accelerator.

\begin{figure}[!ht]
\centering
\includegraphics[trim = 40mm 25mm 25mm 10mm,clip,width=0.83\textwidth]{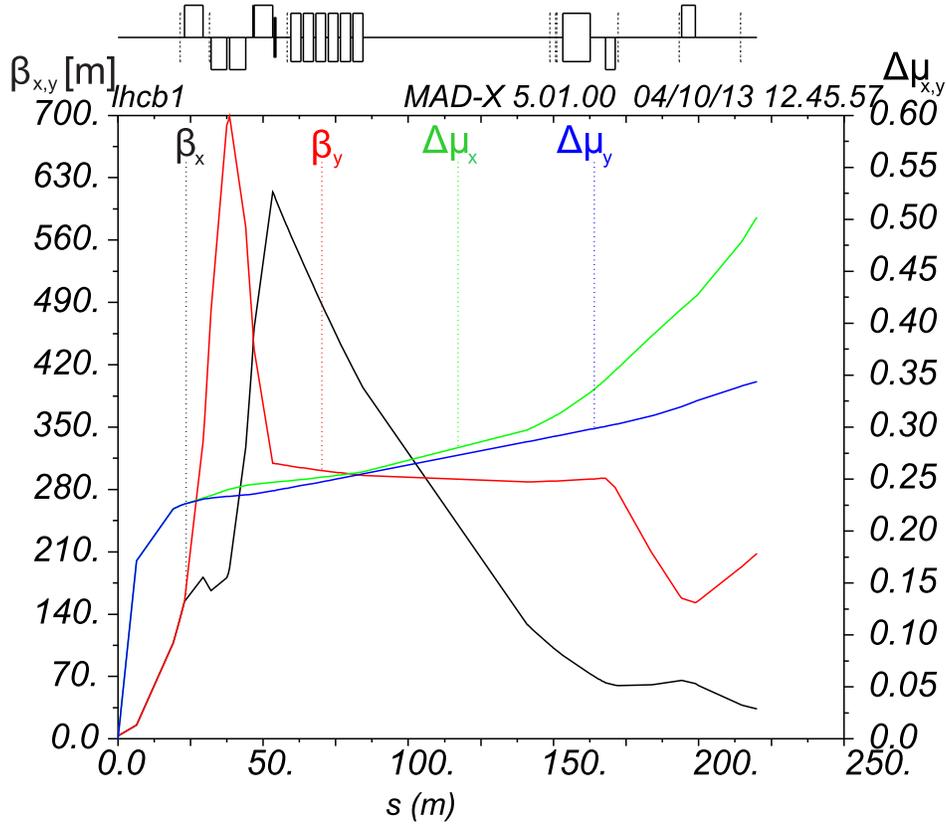}
\caption{(color online) The horizontal $\beta_x$ and vertical betatron amplitude $\beta_y$ for
the LHC $\beta^{*}=3.5$~m optics. The horizontal 
$\Delta\mu_{x}$ and vertical phase advance $\Delta\mu_{y}$ are also shown, 
these functions are normalized to $2\pi$. The plot shows that the horizontal 
phase advance $\Delta\mu_{x}=\pi$ close to the far RP unit.}
\label{betx_bety}
\end{figure}

\begin{figure}[!ht]
\centering
\includegraphics[trim = 40mm 25mm 25mm 10mm, clip, width=0.83\textwidth]{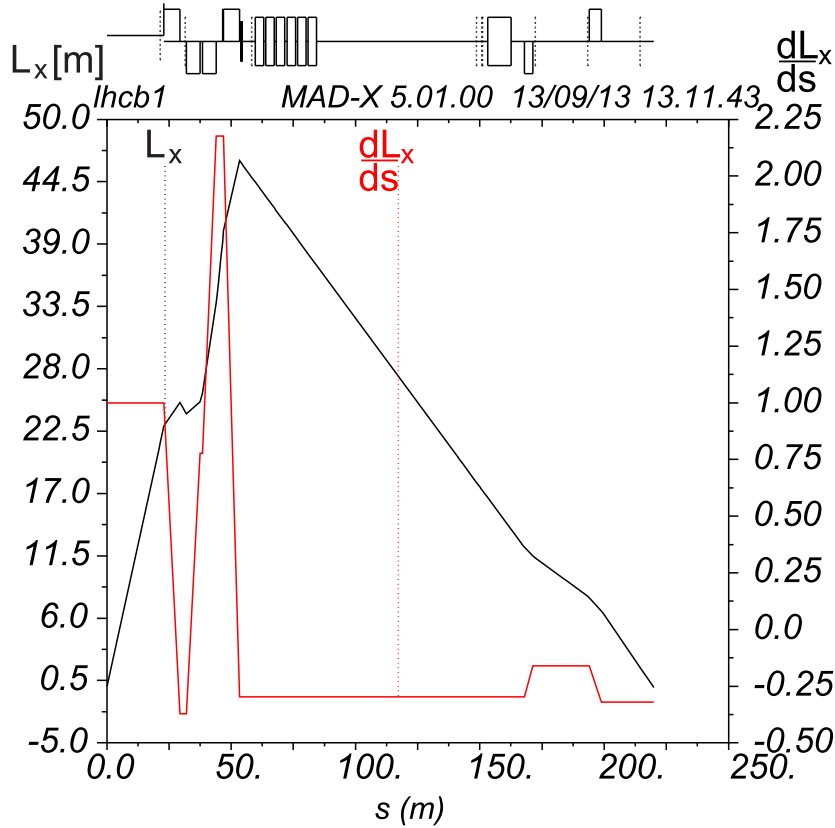}
\caption{(color online) The horizontal effective length $L_x$ and its derivative $\rmd
L_x/\rmd s$ with respect to $s$ as a function of the distance $s$ in case of
the LHC $\beta^{*}=3.5$~m optics. The evolution of the optical functions is
shown starting from IP5 up to the Roman Pot stations. The plot indicates that
$L_{x}=0$ 
close to the far RP unit, thus
in the reconstruction of proton kinematics, 
$\rmd L_{x}/\rmd s$ is used instead of $L_{x}$ 
.}
\label{Lx_dLxds}
\end{figure}

In case of the LHC nominal optics the coupling coefficients are, by design,
equal to zero
\begin{equation}
 	m_{13},...,m_{42}=0\,.
	\label{coupling_coefficients}
\end{equation}
Moreover, for elastically scattered protons the contribution of the vertex
position $(x^{*},y^{*})$ in~\eref{proton_trajectories} is canceled due to the
anti-symmetry of the elastic scattering angles of the two diagonals. Also,
those terms of~\eref{proton_trajectories} which are proportional to the
horizontal or vertical dispersions $D_{x,y}$ vanish, since $\Delta p=0$ for
elastic scattering. Furthermore, the horizontal phase advance
$\Delta\mu_{x}=\pi$ at $219.59$~m, shown in \fref{betx_bety}, and consequently
the horizontal effective length $L_{x}$ vanishes close to the far RP unit, as
it is shown in~\fref{Lx_dLxds}. Therefore, 
$\rmd L_{x}/\rmd s$ is used 
for the reconstruction of the kinematics of proton-proton scattering.

\begin{figure}[!ht]
\centering
\includegraphics[trim = 40mm 25mm 25mm 10mm, clip, width=0.83\textwidth]{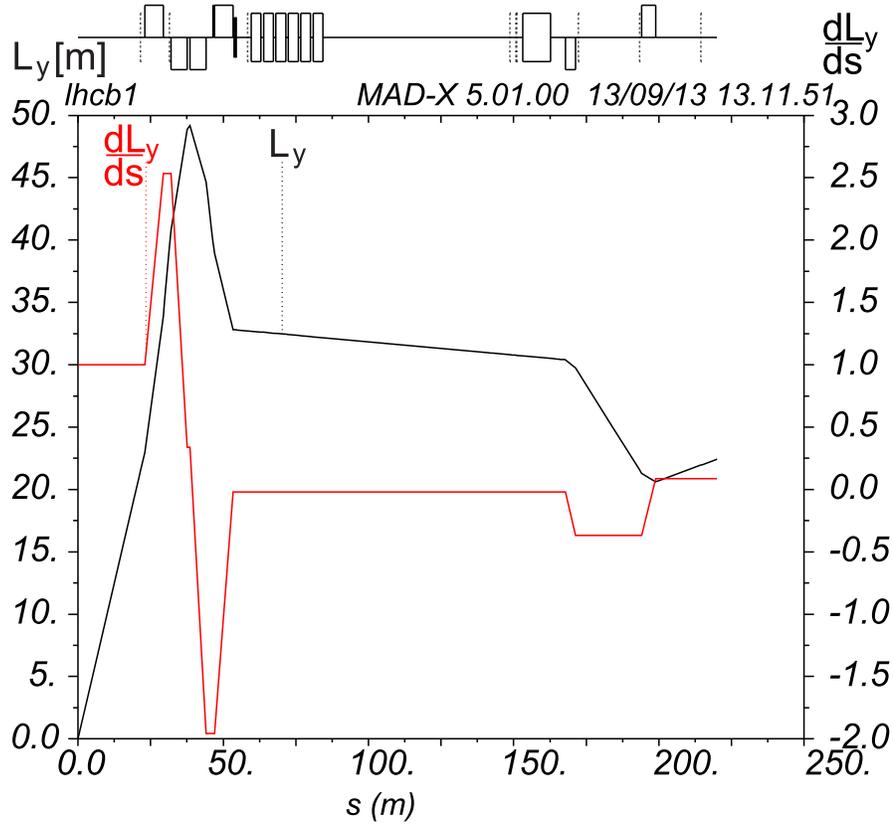}
\caption{(color online) The evolution of the vertical effective length $L_y$ and its
derivative $\rmd L_y/\rmd s$ 
between IP5 and the location of the Roman Pot stations,
for the $\beta^{*}=3.5$~m optics of the LHC.  }
\label{Ly_dLyds}
\end{figure}

\begin{figure}[!ht]
\centering
\includegraphics[trim = 40mm 25mm 27mm 10mm, clip, width=0.83\textwidth]{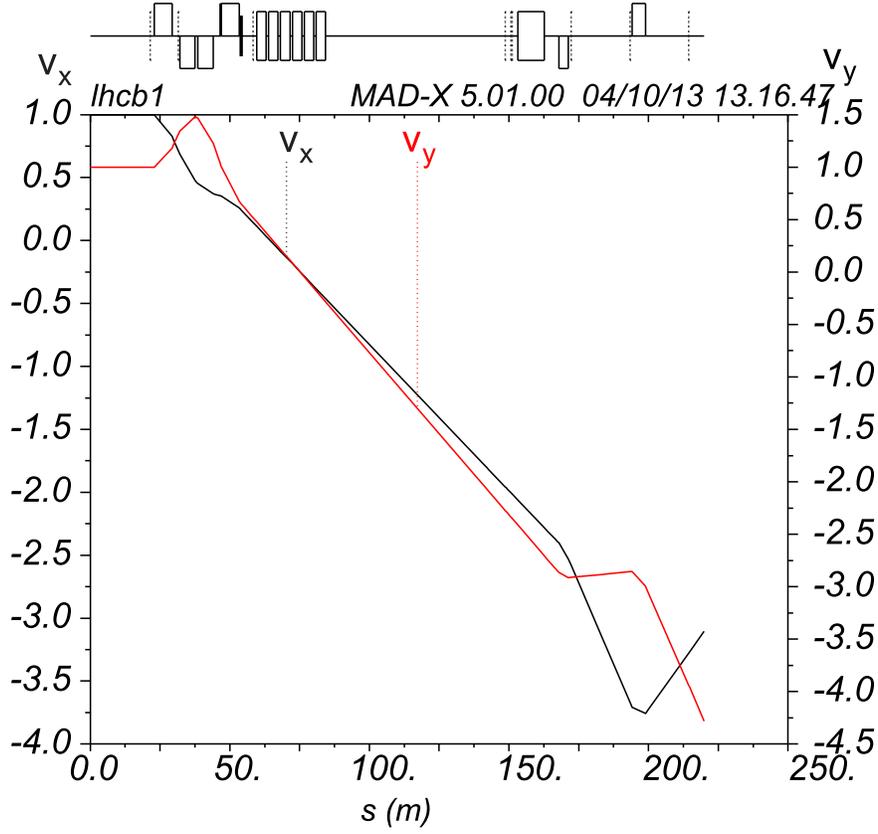}
\caption{(color online) The
evolution of the  horizontal $\nu_x$ and vertical $\nu_y$ magnifications,
for the $\beta^{*}=3.5$~m optics of the LHC.}
	\label{nux_nuy}
\end{figure}

In summary, the kinematics of elastically scattered protons at IP5 can be
reconstructed on the basis of RP proton tracks
using~\eref{proton_trajectories}:
\begin{eqnarray}
	\Theta_{y}^{*} \approx \frac{y}{L_{y}}\,,\,
\qquad
	\Theta_{x}^{*} \approx \frac{1}{\frac{\rmd L_{x}}{\rmd s}}\left(\Theta_{x}-\frac{\rmd v_{x}}{\rmd s}x^{*}\right)\,,\,
\qquad
	x^{*}=\frac{x}{v_x}\,.
	\label{proton_kinematics_reconstruction}
\end{eqnarray}

The vertical effective length $L_{y}$ and the horizontal magnification $v_x$
are applied in~\eref{proton_kinematics_reconstruction} due to their sizeable
values, as shown in figures~\ref{Ly_dLyds} and~\ref{nux_nuy}.  As the values of the
reconstructed angles are inversely proportional to the optical functions, the
errors of the optical functions 
dominate the systematic errors of the 
final, physics results of TOTEM RP measurements.

The proton transport matrix $T\left(s;\mathcal{M}\right)$, calculated with
MAD-X~\cite{MADX}, is defined by the machine settings $\mathcal{M}$, which are
obtained on the basis of several data sources: the magnet currents are first
retrieved from TIMBER \cite{TIMBER} and then converted to magnet strengths with
LSA \cite{LSA}, implementing the conversion curves measured by
FIDEL~\cite{FIDEL}. The WISE database~\cite{WISE} contains the measured
imperfections (field harmonics, magnet displacements and rotations) included in
$\mathcal{M}$.

\section{Machine imperfections}
\label{machine_imperfections}

The real LHC machine~\cite{LHC_machine} is subject to additional imperfections $\Delta \mathcal{M}$, not measured well enough so far, which alter the
transport matrix by $\Delta T$:
\begin{equation}
\label{imperf_mach_eq}
	T\left.(s;\, \mathcal{M}\right) \rightarrow T\left.(s;\, \mathcal{M}+\Delta \mathcal{M}\right) = T\left.(s;\, \mathcal{M}\right)+\Delta T .
\end{equation}
The most important
transport matrix imperfections are due to:
\begin{itemize}
\item[--] the magnet current--strength conversion error: $\sigma(k)/k\approx10^{-3}$,
\item[--] the beam momentum offset: $\sigma(p)/p \approx 10^{-3}\,$.
\end{itemize}
Their impact on the 
 important optical functions 
$L_{y}$ and $\rmd L_{x}/\rmd s$ %
is presented in~\tref{Lysensitivity}.
It is clearly visible that the imperfections of the inner triplet (the so called MQXA and MQXB magnets) are of high influence on the transport matrix while 
the optics is less sensitive to the strength of the quadrupoles MQY and MQML.
 
Other imperfections
that are of lower, but not negligible, significance:
\begin{itemize}
	\item[--] magnet rotations: $\delta\phi\approx1$ mrad,
	\item[--] beam harmonics: $\delta B/B \approx 10^{-4}$,
	\item[--] power converter errors: $\delta I/I \approx 10^{-4}$,
	\item[--] magnet positions: $\delta x,\delta y \approx 100\,\mu$m.
\end{itemize}
Generally, as 
indicated in~\tref{Lysensitivity}, for high-$\beta^*$ optics the magnitude of
$\Delta T$ is sufficiently small from the viewpoint of data analysis.

However, the
sensitivity  of the 
low-$\beta^*$ optics to the machine imperfections is significant
and cannot be neglected. 

\begin{table}[hbt]
\begin{center}
\begin{tabular}{  c | c | c | c | c |} \cline{2-5}
& \multicolumn{2}{|c|}{ } & \multicolumn{2}{|c|}{ }\\[0.01ex] 
& \multicolumn{2}{|c|}{ ${\delta L_{y,b_1,far}/L_{y,b_1,far}}$\,[\%]} & \multicolumn{2}{|c|}{ $\delta\left(\frac{\rmd L_{x,b_1}}{\rmd s}\right)/\frac{\rmd L_{x,b_1}}{\rmd s}$\,[\%]}  \\[1.2ex] \hline
            \multicolumn{1}{|c|}{ Perturbed element}   			& ${\beta^{*}=3.5\,m}$		& ${\beta^{*}=90\,m}$	& ${\beta^{*}=3.5\,m}$		& ${\beta^{*}=90\,m}$	\\ \hline
            \multicolumn{1}{|c|}{MQXA.1R5}        			& $\phantom-0.98$            	& $\phantom-0.14$    	& $-0.46$ 		& $-0.42$ \\
            \multicolumn{1}{|c|}{MQXB.A2R5}       			& $-2.24$            		& $-0.23$    		& $\phantom-0.33$  	& $\phantom-0.31$ \\
           \multicolumn{1}{|c|}{MQXB.B2R5}       			& $-2.42$            		& $-0.25$   		& $\phantom-0.45$  	& $\phantom-0.42$ \\
            \multicolumn{1}{|c|}{MQXA.3R5}        			& $\phantom-1.45$       	& $\phantom-0.20$    	& $-1.14$ 		& $-1.08$ \\
            \multicolumn{1}{|c|}{MQY.4R5.B1}      			& $-0.10$           	  	& $-0.01$    		& $-0.02$ 		& $\phantom-0.00$ \\
            \multicolumn{1}{|c|}{MQML.5R5.B1}     			& $\phantom-0.05$             	& $\phantom-0.04$    	& $\phantom-0.05$ 	& $\phantom-0.06$ \\
            \multicolumn{1}{|c|}{$\Delta$p$_{b_1}$/p$_{b_1}$}   	& $-2.19$             		& $\phantom-0.01$   	& $-0.79$ 		& $\phantom-0.71$ \\ \hline
            \multicolumn{1}{|c|}{Total sensitivity} 			& $\phantom-4.33$   	        & $\phantom-0.42$   	& $\phantom-1.57$	& $\phantom-1.46$ \\ \hline
\end{tabular}
\caption{Sensitivity of the vertical effective length $L_{y,b_{1}}$ and $\rmd
L_{x,b_{1}}/\rmd s$ to 1~$\permil$ deviations of magnet strengths or beam
momentum for low- and high-$\beta^*$ optics of the LHC beam 1. The subscript
$b_{1}$ indicates Beam 1. Only the most important contributions are presented.}
\label{Lysensitivity}
\end{center}
\end{table}
 
The proton reconstruction is based on~\eref{proton_kinematics_reconstruction}.
Thus it is necessary to know the effective lengths $L_{x,y}$ and their
derivatives with an uncertainty better than $1$--$2\,\%$ in order to measure
the total cross-section $\sigma_{\rm tot}$ with the aimed uncertainty of
\cite{TDR}. The currently available $\Delta \beta/\beta$ beating measurement
with an error of $5-10$~\% does not allow to estimate $\Delta T$ with the
uncertainty, required by the TOTEM physics program \cite{LHCOptics}.  However,
as it is shown in the following sections, $\Delta T$ can be determined well
enough from the proton tracks in the Roman Pots, by exploiting the properties
of the optics and those of the elastic $pp$ scattering, 
so that the aimed 1\% relative uncertainty in the determination of the total pp cross-section becomes within the reach of TOTEM. 

\section{Correlations in the transport matrix}
\label{corrInTranspMatrix}

The transport matrix $T$ defining the proton transport from IP5 to the RPs is a product of matrices
that describe the magnetic field of 
the lattice elements along the proton trajectory. 
The imperfections of 
the
individual magnets 
alter the cumulative transport function. It turns out that
independently of the origin of the imperfection (strength of any of the magnets, beam momentum offset) the transport matrix is altered in a similar way, as can
be described quantitatively with eigenvector decomposition, discussed in~\sref{corrMatrixDecomp}. 

\subsection{Correlation matrix of imperfections}
\label{corrMatrixDecomp}
Assuming that the imperfections discussed in~\sref{proton_transport} are independent, the covariance matrix describing
the relations among the errors of the optical functions can be calculated:
\begin{equation}
	V=\mathrm{Cov}(\Delta T_r)={E}\left(\Delta T_r  \Delta T_r^{{T}} \right),
\end{equation}
where $T_r$ is the 
relevant 8-dimensional subset of the transport matrix
\begin{equation}
T_r^{\mbox{\rm \small T}} = (v_x,L_x,\frac{\rmd v_x}{\rmd s},\frac{\rmd L_x}{\rmd s},v_y,L_y,\frac{\rmd v_y}{\rmd s},\frac{\rmd L_y}{\rmd s})\,,
\end{equation}  
which is presented as a vector for simplicity.
 
The optical functions contained in $T_r$ differ by orders of magnitude and, are expressed in different physical units. Therefore, a normalization of $V$ is necessary and the use of the correlation matrix $C$, defined as 
\begin{equation}
	C_{i,j} = \frac{V_{i,j}}{\sqrt{V_{i,i} \cdot V_{j,j}}}\,,
\end{equation}
is preferred.
An identical behaviour of uncertainties for both beams was observed and therefore it is enough to study the Beam 1. In case of the $\beta^*=3.5\,$m optics the following error correlation matrix is obtained:

{\small
\begin{eqnarray} 
\label{covtab3p5}	
C & = & {
\left( \begin{array}{cccccccc}
\phantom-1.00 	& \phantom-0.74 & -0.42 		& -0.80 		& -0.51 		& -0.46 		& -0.61 		& -0.44 \\ 
\phantom-0.74 	& \phantom-1.00 & -0.63 		& -1.00 		& -0.25 		& -0.30 		& -0.32 		& -0.29 \\ 
-0.42 			& -0.63 		& \phantom-1.00 & \phantom-0.62 & \phantom-0.03 & \phantom-0.07 & \phantom-0.01 & \phantom-0.08 \\ 
-0.80 			& -1.00 		& \phantom-0.62 & \phantom-1.00 & \phantom-0.29 & \phantom-0.33 & \phantom-0.37 & \phantom-0.32 \\ 
-0.51 			& -0.25 		& \phantom-0.03 & \phantom-0.29 & \phantom-1.00 & \phantom-0.99 & \phantom-0.98 & \phantom-0.98 \\ 
-0.46 			& -0.30 		& \phantom-0.07 & \phantom-0.33 & \phantom-0.99 & \phantom-1.00 & \phantom-0.96 & \phantom-1.00 \\ 
-0.61 			& -0.32 		& \phantom-0.01 & \phantom-0.37 & \phantom-0.98 & \phantom-0.96 & \phantom-1.00 & \phantom-0.95 \\ 
-0.44 			& -0.29 		& \phantom-0.08 & \phantom-0.32 & \phantom-0.98 & \phantom-1.00 & \phantom-0.95 & \phantom-1.00 \\
\end{array} \right)}\,.
\end{eqnarray}
} 

The non-diagonal elements of $C$, which are close to $\pm1$, indicate strong correlations between the elements of $\Delta T_r$. Consequently, the machine imperfections alter {\it correlated} groups of optical functions.

This observation can be further quantified by the eigenvector decomposition of $C$, which yields the following vector of eigenvalues $\lambda(C)$ for the $\beta^*=3.5\,$m optics:
\begin{equation}
	\lambda(C)=\left(4.9,\,2.3,\,0.53,\,0.27,\,0.01,\,0.01,\,0.00,\,0.00\right).
\end{equation}
Since the two largest eigenvalues $\lambda_{1}=4.9$ and $\lambda_{2}=2.3$ dominate the others, the correlation system is practically two dimensional with
the following two eigenvectors
\begin{eqnarray}
	v_{1}&=\left(\phantom-0.35,\,\phantom-0.30,\,-0.16,\,-0.31,\,-0.40,\,-0.41,\,-0.41,\,-0.40\right)\,, \\
	v_{2}&=\left(-0.26,\, -0.46,\, \phantom-0.47,\, \phantom-0.45,\, -0.29,\, -0.27,\, -0.25,\, -0.28\right)\,.
\end{eqnarray}

Therefore, contributions of the individual lattice imperfections cannot be evaluated. On the other hand, as the imperfections alter approximately only a two-dimensional subspace, a measurement of a small set
of weakly correlated optical functions would theoretically yield an approximate knowledge of $\Delta T_r$. 

\subsection{Error estimation of the method}
\label{methErrEstim}

Let us assume for the moment that we can precisely reconstruct the contributions to $\Delta T_r$ of the two most significant eigenvectors 
while neglecting 
that of the others. The error of such reconstructed transport matrix can be estimated by evaluating the contribution of the remaining eigenvectors:
\begin{equation}
 \delta \Delta T_{r,i}=\sqrt{E_{i,i} \cdot V_{i,i}}\, ,
\end{equation}
where
\begin{equation}
E = N \cdot 
\left( \begin{array}{ccccc}
0 & 0 & 0 & 0 &0 \\
0 & 0 & 0 & 0 &0 \\
0 & 0 & \lambda_3 & 0 &0 \\ 
\vdots &  & & \ddots & \vdots \\
0 & 0 & 0 & 0 & \lambda_8\\
\end{array} \right) \cdot N^{\mathrm{T}} 
\end{equation}
and $N = \left(\nu_1,..., \nu_8 \right)$ is the basis change matrix composed of eigenvectors $\nu_i$ corresponding to the eigenvalues $\lambda_i$. 

The relative optics uncertainty before and after the estimation of the most significant eigenvectors is summarized in~\tref{theor_errors}. 
\begin{table}[hbt]
    \begin{center}
        \begin{tabular}{ c | c | c | c | c |} \cline{2-5}
                                                              					& $v_{x,far}$   & $L_{x,far}$ & $\frac{\rmd v_x}{\rmd s}$ & $\frac{\rmd L_x}{\rmd s}$ \\ \cline{1-5} 
	\multicolumn{1}{|c|}{$T_{r,i}$}                               				& $-3.1$ 		& $-1.32\cdot10^{-1}$\,m 	& $\phantom{-}3.1\cdot10^{-2}$\,m$^{-1}$ 	& $-3.21\cdot10^{-1}$  \\ \cline{1-5}
	\multicolumn{1}{|c|}{\large $\frac{\sqrt{V_{i,i}}}{\left|T_{r,i}\right|}$ [\%]} 	& $2.0\cdot10^{-1}$	& $3.4\cdot10^{2}$       	& $4.2\cdot10^{-1}$    		& $1.6$     \\ 
	\multicolumn{1}{|c|}{\large $\frac{\delta \Delta T_{r,i}}{\left|T_{r,i}\right|}$ [\%]} 	& $9.5\cdot10^{-2}$ 	& $9.1\cdot10^{1}$		& $2.6\cdot10^{-1}$ 		& $3.4\cdot10^{-1}$ \\
	\hline
        \end{tabular}

\vspace{20pt}

        \begin{tabular}{ c | c | c | c | c |} \cline{2-5}
                                                              					& $v_{y,far}$   	& $L_{y,far}$ 		& $\frac{\rmd v_y}{\rmd s}$ 		& $\frac{\rmd Ly}{\rmd s}$ \\ \cline{1-5} 
	\multicolumn{1}{|c|}{$T_{r,i}$}                               				&  $-4.3$ 		& $\phantom{-}2.24\cdot10^{1\phantom{-}}$\,m 			& $-6.1\cdot10^{-2}$\,m$^{-1}$ & $\phantom{-}8.60\cdot10^{-2}$ \\ \cline{1-5}
	\multicolumn{1}{|c|}{\large $\frac{\sqrt{V_{i,i}}}{\left|T_{r,i}\right|}$ [\%]} 	&  $6.8\cdot10^{-1}$	& $4.3$ 		& $5.9\cdot10^{-1}$ 		& $1.5\cdot10^{1}$ \\ 
	\multicolumn{1}{|c|}{\large $\frac{\delta \Delta T_{r,i}}{\left|T_{r,i}\right|}$ [\%]} 	& $6.1\cdot10^{-2}$ 	& $6.4\cdot10^{-1}$ 	& $8.3\cdot10^{-2}$ 		& $2.75$\\
	\hline
        \end{tabular}

        \caption{Nominal values of the optical functions $T_{r,i}$ and their relative uncertainty before ($\sqrt{V_{i,i}}/\left|T_{r,i}\right|$) and after ($\delta \Delta T_{r,i}/\left|T_{r,i}\right|$) the determination of the two most significant eigenvectors ($\beta^*=3.5\,$m, Beam 1).}
	 \label{theor_errors}
    \end{center}
\end{table}
According to the table, even if we limit ourselves only to the first two most significant eigenvalues,
the uncertainty of optical functions due to machine imperfections drops significantly. In particular, in case of $\rmd L_x/\rmd s$ and $L_y$ a
significant error reduction down to a per mil level is observed. Unfortunately, due to $\Delta \mu_{x}=\pi$ (\fref{betx_bety}), the uncertainty of $L_x$, although importantly improved, remains very large and the use of $\rmd L_x/\rmd s$ for proton kinematics reconstruction should be preferred. 

In the following sections a practical numerical method of inferring the optics from the RP proton tracks is presented and its validation with Monte Carlo calculations is reported. 

\section{Optics estimators from proton tracks measured by Roman Pots ($\boldsymbol\beta^{*}$=3.5 m optics)}
\label{constrTracksRP}

The TOTEM experiment can select the elastically scattered protons with high purity and efficiency~\cite{Antchev:2011zz,Antchev:2011vs}. 
The RP detector system, due to its high resolution ($\sigma(x,y) \approx 11\,\mu$m, $\sigma(\Theta_{x,y}) \approx 2.9\,\mu$rad), can measure very precisely the proton angles, positions and the angle-position relations on an event-by-event basis. These quantities can be used to define a set of estimators characterizing
the correlations between the elements of the transport matrix $T$ or between the transport matrices of the two LHC beams. Such a set of estimators $\hat{R}_1, ..., \hat{R}_{10}$ (defined in the next sections) is exploited to reconstruct, for both LHC beams, the imperfect transport matrix $T(\mathcal{M})+\Delta T$ defined in~\eref{imperf_mach_eq}. 

\subsection{Correlations between the beams}

\begin{figure}[!ht]
	\centering
	\includegraphics[width=0.5\textwidth]{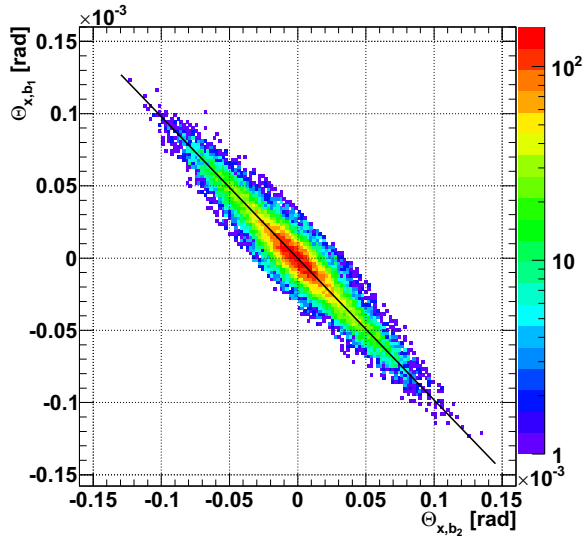}
   	\caption{(color online) Beam 1 and 2 elastic scattering angle correlation in the horizontal plane $(\Theta_{x,b_1}, \Theta_{x,b_2})$ of protons detected by the Roman Pots.}
	\label{ratioofdLxds}
\end{figure}

\begin{figure}[!tbs]
    \centering
    \includegraphics[width=0.5\textwidth]{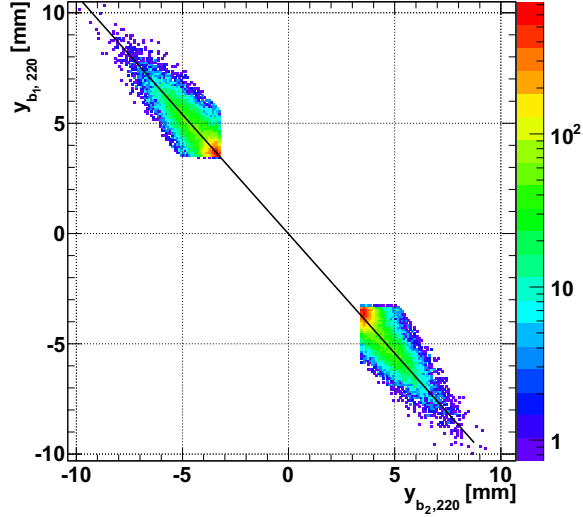}
	\caption{(color online) Correlation between positions (vertical projections) of elastically scattered protons detected in Beam 1 and 2. The sharp edges are due to the vertical acceptance limits of the detectors.}
    \label{ratioofLy}
\end{figure}

Since the momentum of the two LHC beams is identical, the elastically scattered protons will be deflected symmetrically from their nominal trajectories of Beam 1 and Beam 2: 
\begin{eqnarray}
    \Theta^{*}_{x,b_1} = -\Theta^{*}_{x,b_2}\,, 
    \Theta^{*}_{y,b_1} = -\Theta^{*}_{y,b_2}\,, 
    \label{collinearity_cut}
\end{eqnarray}
which allows to compute ratios $R_{1,2}$ relating the effective lengths at the RP locations of the two beams. From~\eref{proton_trajectories} and~\eref{collinearity_cut} we obtain: 
\begin{eqnarray}
    R_{1}&\equiv\frac{\Theta_{x,b_1}}{\Theta_{x,b_2}} \approx \frac{\frac{\rmd L_{x,b_1}}{\rmd s}\cdot\Theta^{*}_{x,b_1}}{\frac{\rmd L_{x,b_2}}{\rmd s}\cdot\Theta^{*}_{x,b_2}} =- 
        \frac{\frac{\rmd L_{x,b_1}}{\rmd s}}{\frac{\rmd L_{x,b_2}}{\rmd s}}\,,
    \label{ratiodLxds}
\end{eqnarray}
\begin{eqnarray}
    R_{2}&\equiv\frac{y_{b_1,far}}{y_{b_2,far}} \approx -\frac{L_{y,b_1,far}}{L_{y,b_2,far}} \,,
    \label{ratioLy}
\end{eqnarray}
where the subscripts $b_1$ and $b_2$ indicate Beam 1 and 2, respectively. Approximations present in~\eref{ratiodLxds} and~\eref{ratioLy} represent the impact of statistical effects such as detector resolution, beam divergence and primary vertex position distribution. 
The estimators $\hat{R}_1$ and $\hat{R}_2$ are finally obtained from the $(\Theta_{x,b_1}, \Theta_{x,b_2})$ and $(y_{b_1,far},y_{b_2,far})$ distributions and are defined with the help of the distributions' principal eigenvector,
as illustrated in figures~\ref{ratioofdLxds} and~\ref{ratioofLy}. 

The width of the distributions is determined by the beam divergence and the vertex contribution, which leads to 0.5\% uncertainty on the eigenvector's slope parameter.

\subsection{Single beam correlations}
\begin{figure}[htbs!]
    \centering
    \includegraphics[width=0.5\textwidth]{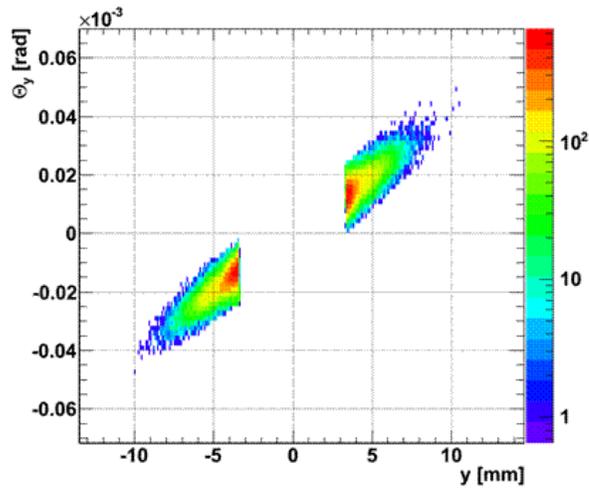}
    \caption{(color online) Correlation between vertical position and angle of elastically scattered protons at the RP of Beam 1.}
    \label{ratioofdLydsLy}
\end{figure}

The distributions of proton angles and positions measured by the Roman Pots define the ratios of 
certain elements of the transport matrix $T$, defined by~\eref{proton_trajectories} and~\eref{transport_matrix}. First of all, $\rmd L_y/\rmd s$ and $L_y$ are related by 
\begin{eqnarray}
	R_3\equiv\frac{\Theta_{y,b_1}}{y_{b_1}} \approx \frac{\frac{\rmd L_{y,b_1}}{\rmd s}}{L_{y,b_1}}\,,\;\qquad \quad
	R_4\equiv\frac{\Theta_{y,b_2}}{y_{b_2}} \approx \frac{\frac{\rmd L_{y,b_2}}{\rmd s}}{L_{y,b_2}}\,.
\end{eqnarray} 
The corresponding estimators $\hat{R}_3$ and $\hat{R}_4$ can be calculated with an uncertainty of 0.5\% from the distributions as presented in~\fref{ratioofdLydsLy}. 

\begin{figure}[htb!]
    \centering
    \includegraphics[width=0.5\textwidth]{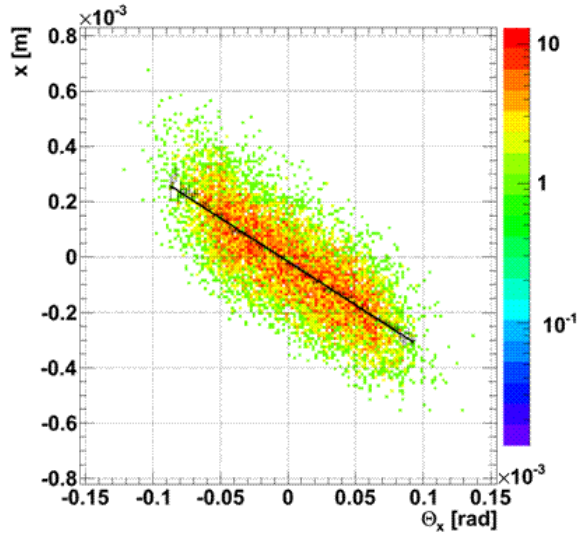}
    \caption{(color online) Correlation between the horizontal angle and position of elastically scattered protons at the RP of Beam 1.}
    \label{ratioofLxdLxds214p4}
\end{figure}

Similarly, we exploit the horizontal dependencies to quantify the relations between $\rmd L_x/\rmd s$ and $L_x$. As $L_x$ is close to $0$, see~\fref{Lx_dLxds},
instead of defining the ratio we rather estimate the position $s_{0}$ along the beam line (with the uncertainty of about $1\,$m), for which $L_x=0$. This is accomplished by resolving 
\begin{eqnarray}
	\frac{L_x(s_{0})}{\rmd L_x(s_1)/\rmd s}= \frac{L_x(s_1)}{\rmd L_x(s_1)/\rmd s} + \left(s_{0}-s_{1}\right)=0\,,
\end{eqnarray}
for $s_{0}$, where $s_1$ denotes the coordinate of the Roman Pot station along the beam with respect to IP5. Obviously, $\rmd L_x(s)/\rmd s$ is constant along the RP station as no magnetic fields are
present at the RP location. The ratios $L_{x}(s_1)/\frac{\rmd L_{x}(s_1)}{\rmd s}$ for Beam 1 and 2, similarly to the vertical constraints $R_3$ and $R_4$, are defined by the proton tracks:
\begin{eqnarray}
    \frac{L_{x}}{\frac{\rmd L_{x}}{\rmd s}}\approx\frac{x}{\Theta_{x}}\,, 
\end{eqnarray}
which is illustrated in~\fref{ratioofLxdLxds214p4}.
In this way two further constraints and the corresponding estimators (for Beam 1 and 2) are obtained:
\begin{equation}
 R_5 \equiv s_{b_1}\mathrm{\,\, and \,\,} R_6 \equiv s_{b_2}\,.
\end{equation}

\subsection{Coupling / rotation}
\label{coupling_rotation}
\begin{figure}[!ht]
    \centering
    \includegraphics[width=0.5\textwidth]{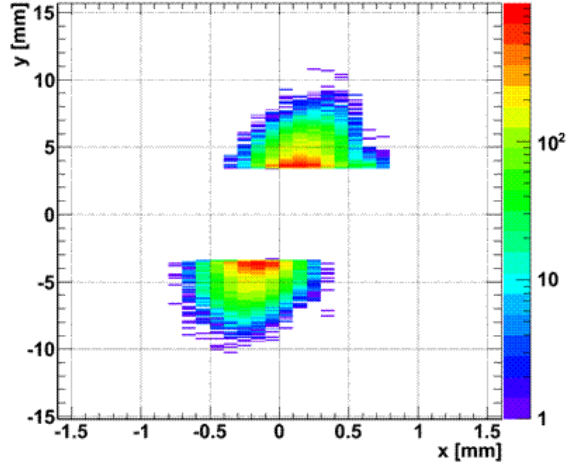}
    \caption{(color online) Vertical vs. horizontal track position at the RP$_{\mathrm{near}}$ of the LHC Beam 1.}
    \label{ratioofre14Ly}
\end{figure}

In reality the coupling coefficients $m_{13},...,m_{42}$ cannot be always neglected, as it is assumed by~\eref{coupling_coefficients}. RP proton tracks can help to determine the coupling components of the
transport matrix $T$ as well, where it is especially important that $L_{x}$ is close to zero at the RP locations. Always based on~\eref{proton_trajectories} and~\eref{transport_matrix},
four additional constraints (for each of the two LHC beams and for each unit of the RP station) can be defined:
\begin{eqnarray}
	R_{7,...,10}\equiv\frac{x_{{near (far)}}}{y_{{near (far)}}} \approx \frac{m_{14,{near (far)}}}{L_{y,{near (far)}}}\,.
	\label{coupling_rotations}	
\end{eqnarray}
The subscripts ``near'' and ``far'' indicate the position of the RP along the
beam with respect to the IP. Geometrically $R_{7,...,10}$ describe the rotation
of the RP scoring plane about the beam axis. Analogously to the previous
sections, 
the estimators $\hat{R}_{7,...,10}$ are obtained from track
distributions as presented in~\fref{ratioofre14Ly} and an uncertainty of $3\%$
is achieved. 

\section{Optical functions estimation}
\label{opticsMatching}

The machine imperfections $\Delta \mathcal{M}$, leading to the transport matrix change $\Delta T$, are in practice determined with the $\chi^2$ minimization procedure:
\begin{equation}
 \widehat{\Delta \mathcal{M}} = arg\,min(\chi^2)\,,
\end{equation}
defined on the basis of the estimators $\hat{R}_1...\hat{R}_{10}\,$, where the {$arg\,min$} function gives the phase space position where the $\chi^2$ is minimized. As it was discussed in~\sref{corrMatrixDecomp}, although the overall alteration of the transport matrix $\Delta T$ can be determined
precisely based on a few optical functions' measurements, the contributions of individual imperfections cannot be established. In terms of optimization, such a problem has no unique solution and additional constraints,
defined by the machine tolerance, have to be added.

Therefore, the $\chi^2$ function is composed of the part defined by the Roman Pot tracks' distributions and the one reflecting the LHC tolerances:
\begin{eqnarray}
\chi^2 = \chi_{Design}^2 + \chi_{Measured}^2\,.
	\label{chi2_constructions}
\end{eqnarray}
The design part
\begin{equation}
\fl\chi_{Design}^2  =  \sum_{i=1}^{12}\left(\frac{k_{i}-k_{i,{\mbox{\tiny MAD-X}}}}
                                                         {\sigma(k_i)}\right)^2 + 
                         \sum_{i=1}^{12}\left(\frac{\phi_{i}-\phi_{i,{\mbox{\tiny MAD-X}}}}
                                                         {\sigma(\phi_i)}\right)^2 +
			 \sum_{i=1}^{2}\left(\frac{p_{i}-p_{i,{\mbox{\tiny MAD-X}}}}
							{\sigma(p_i)}\right)^2\,
				\label{chi2_design}
\end{equation}
where $k_{i}$ and $\phi_{i}$ are the nominal strength and rotation of the $i$th magnet, respectively. Thus~\eref{chi2_design} defines the nominal machine $(k_i,\,\phi_i,\,p_i)$ as an attractor in the phase space.
Both LHC beams are treated simultaneously. Only the relevant subset of machine imperfections $\Delta \mathcal{M}$ was selected. The obtained 26-dimensional 
optimization phase space includes the magnet strengths (12 variables), rotations (12 variables) and beam momentum offsets (2 variables). Magnet rotations are 
included into the phase space, otherwise only the coupling coefficients $m_{13},...,m_{42}$ could induce rotations in the $(x,y)$ plane~\eref{coupling_rotations},
which could bias the result.

The measured part
\begin{equation}
    \chi^2_{Measured} = \sum_{i=1}^{10}\left(\frac{\hat{R}_{i}-R_{i,{\mbox{\tiny MAD-X}}}}
                       {\sigma(\hat{R}_i)}\right)^2
\end{equation} 
contains the track-based estimators $\hat{R}_1...\hat{R}_{10}$ (discussed in detail in~\sref{constrTracksRP}) together with their uncertainty. The subscript ``MAD-X" defines the corresponding values evaluated
with the MAD-X software during the $\chi^2$ minimization.

\Tref{matching_result} presents the results of the optimization procedure for the $\beta^*=3.5\,$m optics used by LHC in October 2010 at beam energy $E=3.5\,$TeV.

The obtained value of the effective length $L_y$ of Beam 1 
is close to the nominal one, while Beam 2 shows a significant change. The same pattern applies to the values of $\rmd L_x/\rmd s$. The error estimation of the procedure is discussed in~\sref{monteCarloTests}.
\begin{table}[H]\renewcommand{\arraystretch}{1.0}\addtolength{\tabcolsep}{-4pt}
\begin{center}
	\begin{tabular}{ c |c|c|c|c|}\cline{2-5}
							&	${L_{y,b_{1},far}}$[m]  	&	${\rmd L_{x,b_{1}}/\rmd s}$	& { ${L_{y,b_2,far}}$[m]} 				&      ${\rmd L_{x,b_2}/\rmd s}$			\\\hline		
	\multicolumn{1}{|c|}{ Nominal}			& 	$22.4$ 				&	$-3.21\cdot10^{-1}$ 		& $18.4$ 						&	$-3.29\cdot10^{-1}$ 	\\\hline
	\multicolumn{1}{|c|}{ Estimated}		&	$22.6$				& 	$-3.12\cdot10^{-1}$ 		& $20.7$						& 	$-3.15\cdot10^{-1}$	\\\hline
	\end{tabular}
\end{center}
	\caption{Selected optical functions of both LHC beams for the $\beta^*=3.5\,$m optics, obtained with the estimation procedure, compared to their nominal values.}
	\label{matching_result}
\end{table}
\vspace{-15pt}
\section{Monte Carlo validation}
\label{monteCarloTests}
In order to demonstrate that the proposed $\hat{R}_i$ optics estimators are effective the method was validated with Monte Carlo simulations.

\begin{figure}[H]
	\begin{center}
		\includegraphics[width=0.49\textwidth]{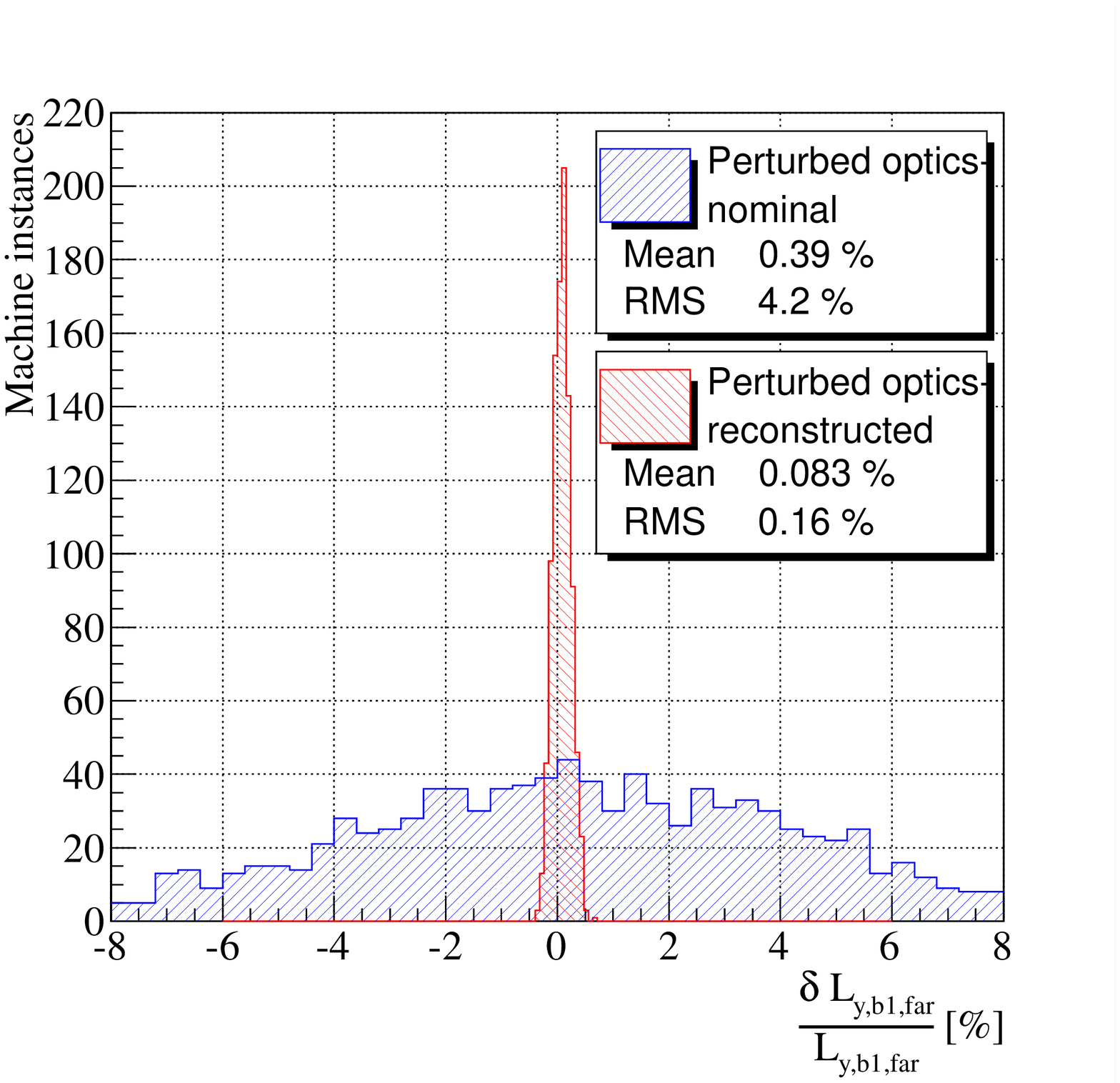}
		\includegraphics[width=0.49\textwidth]{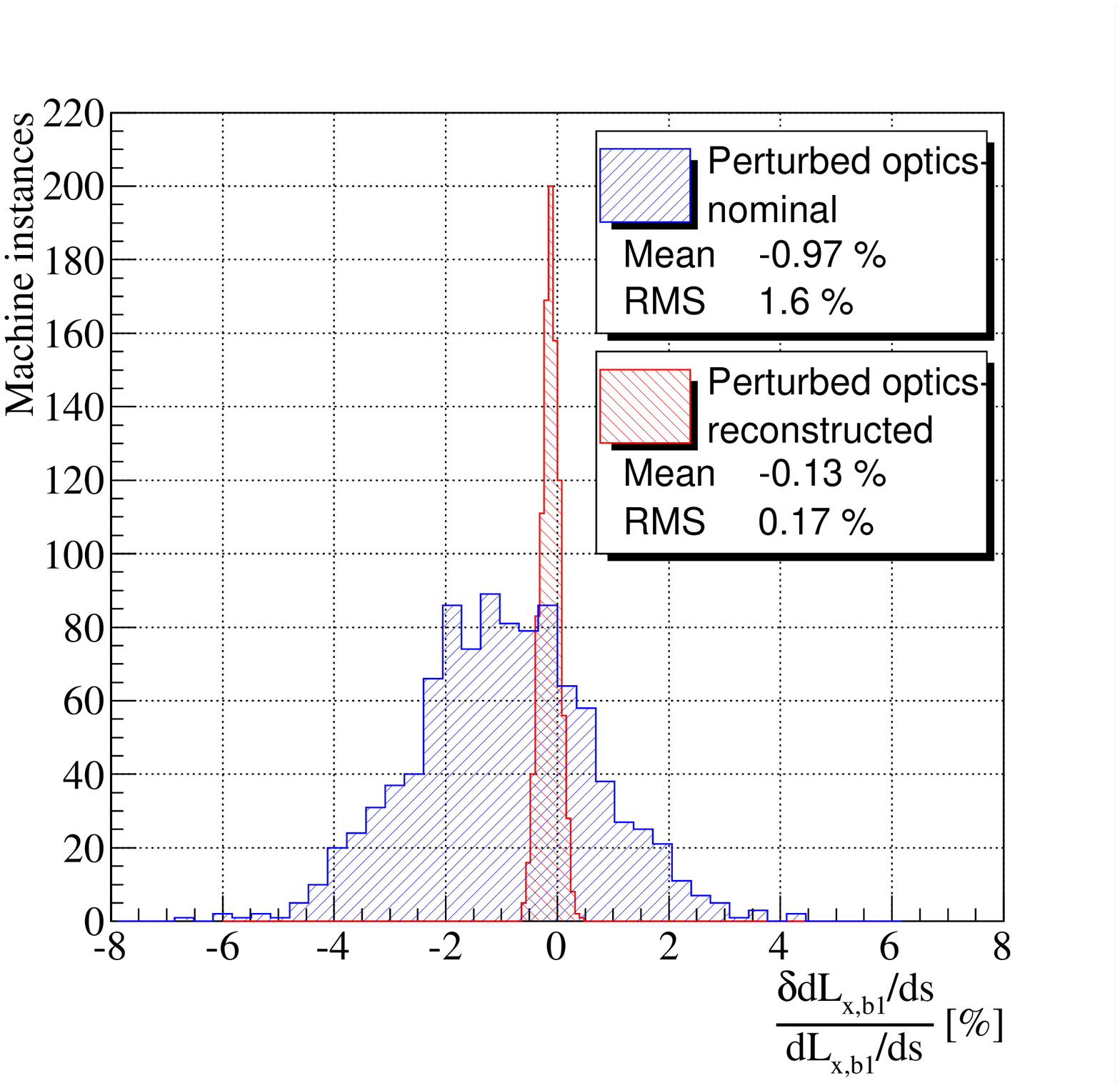}
		\vspace{-10pt}
		\caption{(color online) The MC error distribution of $\beta^{*}=3.5$ m optical functions $L_{y}$ and $\rmd L_{x}/\rmd s$ for Beam 1 at $E=3.5$~TeV, before and after optics estimation. }
		\label{MCresultLy_3_5}
	\end{center}
\end{figure}

\begin{figure}[H]
	\begin{center}
		\includegraphics[trim = 0mm 0mm 0mm 0.1mm, clip, width=0.49\textwidth]{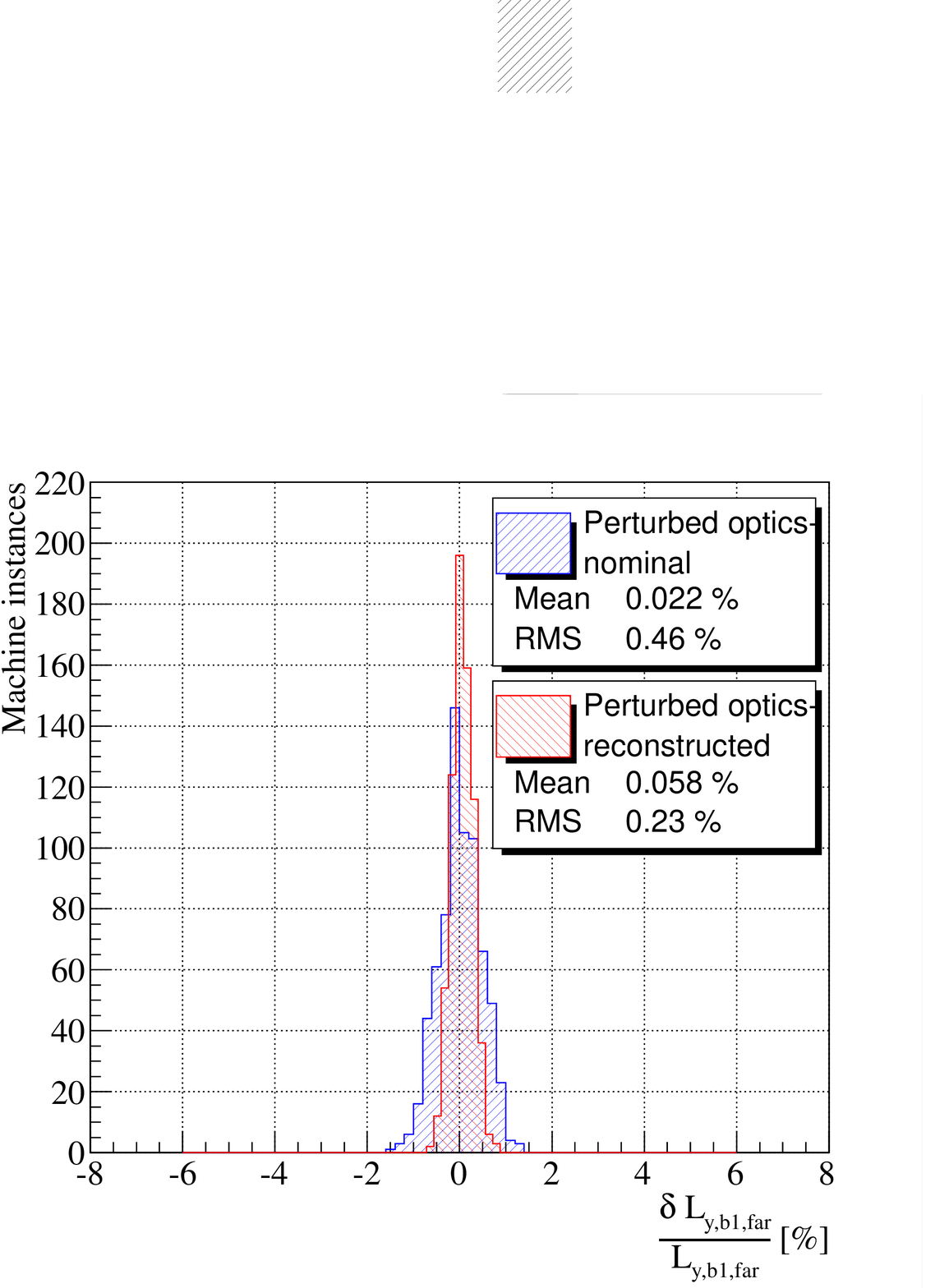}
		\includegraphics[width=0.49\textwidth]{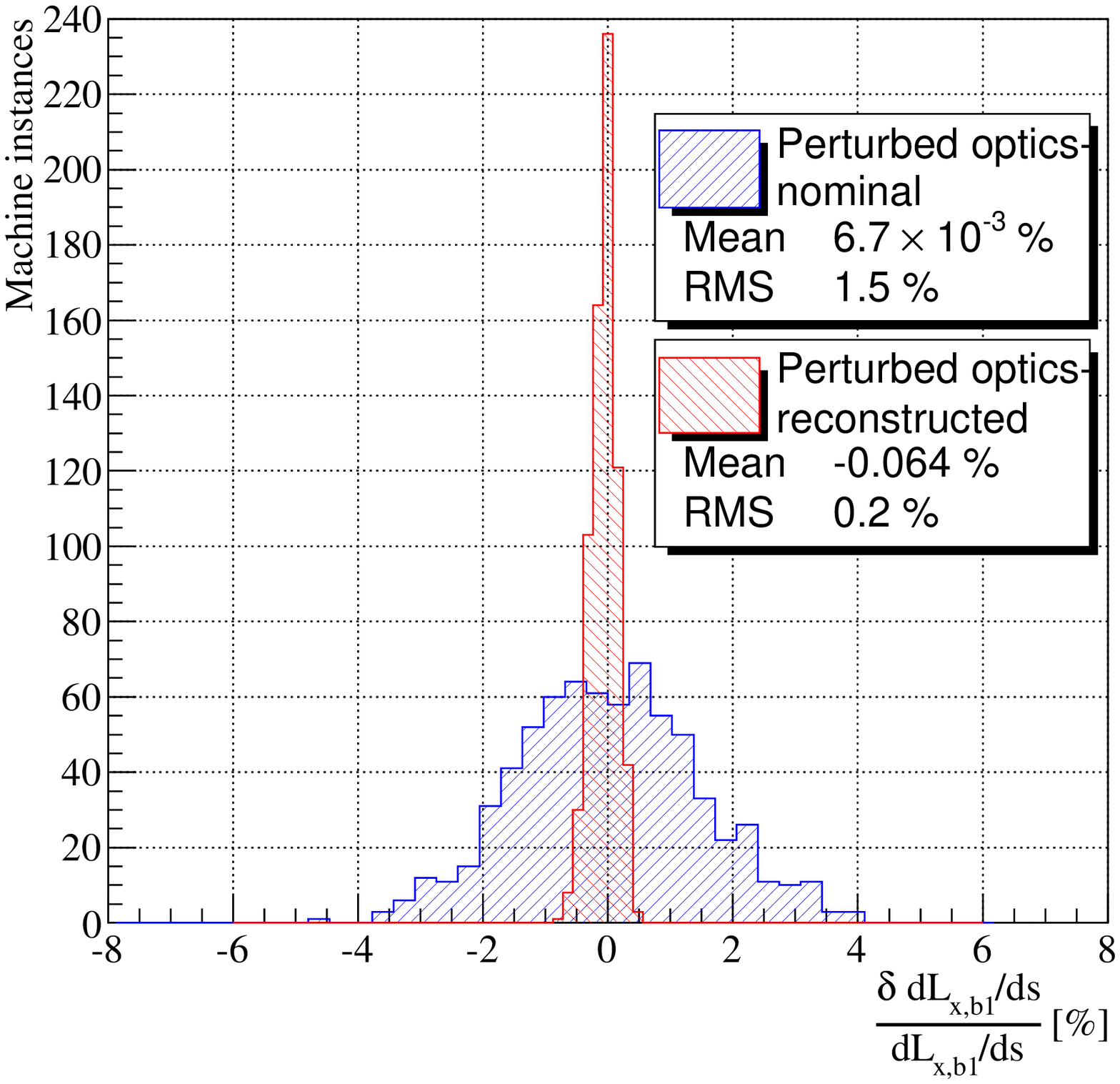}
		\caption{(color online) The MC error distribution of $\beta^{*}=90$ m optical functions $L_{y}$ and $\rmd L_{x}/\rmd s$ for Beam 1 at $E=4$~TeV, before and after optics estimation. }
		\label{MCresultLy_90}
	\end{center}
\end{figure}

In each Monte Carlo simulation the nominal machine settings $\mathcal{M}$ were altered with simulated machine imperfections $\Delta \mathcal{M}$ within their tolerances using Gaussian distributions.
The simulated elastic proton tracks were used afterwards to calculate the estimators $\hat{R}_1...\hat{R}_{10}$. The
study included the impact of 
	\begin{itemize}
		\item[--] magnet strengths, 
		\item[--] beam momenta,
		\item[--] magnet displacements, rotations and harmonics,
		\item[--] settings of kickers,
		\item[--] measured proton angular distribution.
	\end{itemize}
The error distributions of the optical functions $\Delta T$ obtained for $\beta^*=3.5\,$m and $E=3.5\,$TeV are presented in~\fref{MCresultLy_3_5} and~\tref{MCestimations}, while the $\beta^*=90\,$m results
at $E=4\,$TeV are shown in~\fref{MCresultLy_90} and~\tref{MCestimations_90_meters}.

\begin{table}[htb!]\renewcommand{\arraystretch}{1.0}\addtolength{\tabcolsep}{-4pt}
	\begin{center}
   \begin{tabular}{  c | c | c | c | c | c | c |} \cline{2-5}
       & \multicolumn{2}{|c|}{Simulated} & \multicolumn{2}{|c|}{Reconstructed} \\
       & \multicolumn{2}{|c|}{optics distribution} & \multicolumn{2}{|c|}{optics error} \\ \hline
       \multicolumn{1}{|c|}{ Relative optics }                                      & { Mean}   & { RMS}        & { Mean}                   & { RMS}   \\
       \multicolumn{1}{|c|}{distribution}                                           &    [\%]      &     [\%]       &     [\%]                      & [\%]  \\ \hline
       \multicolumn{1}{|c|}{ $\frac{\delta L_{y,b_{1},far}}{L_{y,b_{1},far}}\;$ }               &  $\phantom-0.39$         & $4.2$           & $\phantom-8.3\cdot10^{-2}$    & $0.16$ \\
       \multicolumn{1}{|c|}{ $\frac{\delta \rmd L_{x,b_{1}}/\rmd s}{\rmd L_{x,b_{1}}/\rmd s}\;$ }       &  $-0.97$          & $1.6$           & $-0.13$             & $0.17$  \\\hline
        \multicolumn{1}{|c|}{ $\frac{\delta L_{y,b_2,far} }{L_{y,b_2,far}} \;$ }            &  $-0.14$          & $4.9$           & $\phantom-0.21$    & $0.16$ \\
        \multicolumn{1}{|c|}{ $\frac{\delta \rmd L_{x,b_2}/\rmd s}{\rmd L_{x,b_2}/\rmd s}\;$ }      & $\phantom-0.10$         & $1.7$           & $-9.7\cdot10^{-2}$              & $0.17$  \\\hline

   \end{tabular}
   \caption{The Monte-Carlo study of the impact of the LHC imperfections $\Delta \mathcal{M}$ on selected transport matrix elements $\rmd L_{x}/\rmd s$ and $L_{y}$ for $\beta^*=3.5\,$m at $E=3.5$~TeV. The LHC parameters were altered within their tolerances. The relative errors of $\rmd L_{x}/\rmd s$ and $L_{y}$ (mean value and RMS) characterize the optics uncertainty before and after optics estimation.
   \label{MCestimations} 
} 
\end{center}
\end{table}

\begin{table}[htb!]\renewcommand{\arraystretch}{1.0}\addtolength{\tabcolsep}{-4pt}
	\begin{center}
   \begin{tabular}{  c | c | c | c | c | c | c |} \cline{2-5}
       & \multicolumn{2}{|c|}{Simulated} & \multicolumn{2}{|c|}{Reconstructed} \\
       & \multicolumn{2}{|c|}{optics distribution} & \multicolumn{2}{|c|}{optics error} \\ \hline
       \multicolumn{1}{|c|}{ Relative optics }                                      & { Mean}   		& { RMS}        	& { Mean}                   	& { RMS}   \\
       \multicolumn{1}{|c|}{distribution}                                           &    [\%]      		&     [\%]       	&     [\%]                      & [\%]  \\ \hline
       \multicolumn{1}{|c|}{ $\frac{\delta L_{y,b_{1},far}}{L_{y,b_{1},far}}\;$ }   &  $2.2\cdot10^{-2}$       	& $0.46$           	& $\phantom-5.8\cdot10^{-2}$    & $0.23$ \\
       \multicolumn{1}{|c|}{ $\frac{\delta \rmd L_{x,b_{1}}/\rmd s}{\rmd L_{x,b_{1}}/\rmd s}\;$ }   &  $6.7\cdot10^{-3}$	& $1.5$           	& $-6.4\cdot10^{-2}$            & $0.20$  \\\hline
        \multicolumn{1}{|c|}{ $\frac{\delta L_{y,b_2,far} }{L_{y,b_2,far}} \;$ }    &  $-5\cdot10^{-3}$       	& $0.47$          	& $5.8\cdot10^{-2}$   		& $0.23$ \\
        \multicolumn{1}{|c|}{ $\frac{\delta \rmd L_{x,b_2}/\rmd s}{\rmd L_{x,b_2}/\rmd s}\;$ }      &  $1.8\cdot10^{-2}$       	& $1.5$           	& $-7\cdot10^{-2}$            	& $0.21$  \\\hline

   \end{tabular}
   \caption{The Monte-Carlo study of the impact of the LHC imperfections $\Delta \mathcal{M}$ on selected transport matrix elements $\rmd L_{x}/\rmd s$ and $L_{y}$ for $\beta^*=90\,$m at $E=4$~TeV. The LHC parameters were altered within their tolerances. The relative errors of $\rmd L_{x}/\rmd s$ and $L_{y}$ (mean value and RMS) characterize the optics uncertainty before and after optics estimation.
   \label{MCestimations_90_meters} 
} 
\end{center}
\end{table}

First of all, the impact of the machine imperfections $\Delta \mathcal{M}$ on the transport matrix $\Delta T$, as shown by the MC study, is identical to the theoretical prediction presented in~\tref{theor_errors}. The bias
of the simulated optics distributions is due to magnetic field harmonics as reported by the LHC imperfections database~\cite{WISE}. The final value of mean after optics estimation procedure contributes to the total
uncertainty of the method.

The errors of the reconstructed optical functions are significantly smaller
than evaluated theoretically in~\sref{methErrEstim}. This results from the
larger number  of
design and measured constraints~\eref{chi2_constructions}, employed in the numerical estimation
procedure of~\sref{opticsMatching}. In particular, the collinearity of
elastically scattered protons was exploited in addition. Finally, the achieved
uncertainties of $\rmd L_{x}/\rmd s$ and $L_{y}$ are both lower than
$2.5\,\permil$ for both beams. 

\section{Conclusions}
TOTEM has proposed a novel approach to estimate the optics at LHC. The method,
based on the correlations of the transport matrix, consists 
of the
determination of the optical functions, which are strongly correlated to
measurable combinations of the transport matrix elements.

At low-$\beta^*$ LHC optics, where machine imperfections are more significant,
the method allowed us  to determine the real optics with a per mil level
uncertainty, and also permitted  to assess the errors of the transport matrix errors from the
tolerances of various machine parameters. In case of high-$\beta^*$ LHC optics,
where the machine imperfections have smaller effect on the optical functions,
the method remains effective and reduces the uncertainties to the desired per
mil level.  The method has been validated with the Monte Carlo studies both for
high- and low-$\beta^*$ optics and was successfully used in the TOTEM
experiment 
to calibrate the optics of the LHC accelerator directly from data in physics runs
for precision TOTEM measurements of the total pp cross-section.

\section*{Acknowledgments}
This work was supported by the institutions listed on the front page and partially also by NSF (US), the Magnus Ehrnrooth foundation (Finland), the Waldemar von Frenckell foundation (Finland), the Academy of Finland, the Finnish Academy of Science and Letters (The Vilho, Yrj\"{o} and Kalle V\"{a}is\"{a}l\"{a} Fund), the OTKA grant NK 101438 (Hungary) and the Ch. Simonyi Fund (Hungary).

\begin{flushleft}
\bigskip
Individual members of the TOTEM Collaboration are also from:\\
{$^{\rm a}$ INRNE-BAS, Institute for Nuclear Research and Nuclear Energy, Bulgarian Academy of Sciences, Sofia, Bulgaria,} \\
{$^{\rm b}$ Department of Atomic Physics, E\"otv\"os University,  Budapest, Hungary,}\\
{$^{\rm c}$ Ioffe Physical - Technical Institute of Russian Academy of Sciences, St.Petersburg, Russia,}\\
{$^{\rm d}$ Warsaw University of Technology, Warsaw, Poland,}\\
{$^{\rm e}$ Institute of Nuclear Physics, Polish Academy of Science, Cracow, Poland,}\\
{$^{\rm f}$ SLAC National Accelerator Laboratory, Stanford CA, USA,}\\
{$^{\rm g}$ Penn State University, Dept.~of Physics, University Park, PA USA.}
\bigskip
\end{flushleft}
\vfill\eject

\end{document}